\documentclass[twocolumn,twoside,slac_two]{revtex4}
\usepackage{graphicx}
\usepackage{fancyhdr}
\begin{document}

\title{Interpretation of blazar SEDs based on broad band quasi-simultaneous observations}

%

\author{S.~Rain\`o}
\affiliation{Universit\`a di Bari and INFN Sezione di Bari}
\author{E.~Cavazzuti}
\affiliation{Agenzia Spaziale Italiana (ASI) Science Data Center, I-00044 Frascati (Roma), Italy}

\author{S.~Colafrancesco}
\affiliation{Agenzia Spaziale Italiana (ASI) Science Data Center, I-00044 Frascati (Roma), Italy}

\author{P.~Giommi}
\affiliation{Agenzia Spaziale Italiana (ASI) Science Data Center, I-00044 Frascati (Roma), Italy}

\author{A.~Tramacere}
\affiliation{ISDC, Data Centre for Astrophysics, Chemin d'Ecogia 16, CH1290 Versoix, Switzerland}

\author{on behalf of the \emph{Fermi} LAT Collaboration}

\begin{abstract}

We study the quasi-simultaneous Spectral Energy Distributions (SED) of 48 bright blazars, collected within three months  of the Fermi LAT Bright AGN Sample (LBAS) data taking period, combining Fermi and Swift data with radio NIR-Optical  and hard-X/$\gamma$-ray data.
We have used these SEDs to characterize the peak  position and intensity of both the low and the high-energy features of blazar spectra.
The results have been used to derive empirical relationships that estimate the position of the two peaks from the broad-band colors (i.e. the radio to optical, $\alpha_{ro}$, and optical to X-ray, $\alpha_{ox}$, spectral slopes) and from the $\gamma$-ray spectral index. Our data show that the synchrotron peak frequency is positioned
between 10$^{12.5}$ and 10$^{14.5}$~Hz in broad-lined FSRQs and between 10$~{13}$ and 10$^{17}$~Hz
in featureless BL Lacertae objects. We find that the $\gamma$-ray spectral slope is strongly
correlated with the synchrotron peak energy and with the X-ray spectral index, as
expected at first order in synchrotron - inverse Compton scenarios.
However, simple homogeneous, one-zone, Synchrotron Self Compton (SSC) models cannot explain most of our SEDs, especially in the case of FSRQs and low energy peaked (LBL) BL Lacs. More complex models involving External Compton Radiation or multiple SSC components are required to reproduce the overall SEDs and the observed spectral variability.

\end{abstract}

\maketitle

\thispagestyle{fancy}


\section{Introduction}

The \emph{Fermi Gamma-Ray Space Telescope} was launched on June 11, 2008 from Cape Kennedy, Florida opening a new era in the knowledge of high-energy $\gamma$-ray band. The high sensitivity, large dinamic range and excellent time coverage of the Large Area Telescope, the primary instrument on board \emph{Fermi}, all represent significant advantages over the previous $\gamma$-ray observations and will play a key role in the understanding of the physical processes underlying the high energy emission of blazars. 
Typical AGN are radio-quiet and are found in large numbers of optical and X-ray frequencies.
Blazars are rare extra-galactic objects, about 10$\%$ of the entire population of AGN, where a jet of plasma, moving at relativistic speed is pointing close to the line of sight of the observer \cite{Blandford1978}.  However, blazars strong non-thermal emission over the entire electro-magnetic spectrum makes them the dominant type of extragalactic sources in those energy windows where the accretion onto a supermassive black hole, or other thermal machenism, do not produce much radiation. Although blazars have been observed for tens of years now, the exististing experimental data do not allow one to unambiguously identify the physical mechanism responsible for the electromagnetic emission.
Given the existing high sensitivity detectors to study the low energy component (radio, IR, optical and X-rays), one of the main reasons for this lack of understanding was the low instrumental background between 0.5~MeV to 300~GeV. This poor sensitivity often precluded detailed cross-correlation studies between low and high energy emission models, implying more and higher quality data at $\gamma$-ray energies to better understand these objects. Simultaneous and detailed Spectral Energy Distributions (SEDs) are indeed needed to cover the greatest possible band of the electromagnetic spectrum. In fact, one of the most effective ways of studying blazar's physical properties is through the use of multi-frequency data that, so far, have been carried out almost exclusively on the occasion of large flaring events of a few bright and well known blazars.

In this contribution I report on the properties of broad-band (radio to high energy $\gamma$-ray) SEDs of a sample of 48 \emph{Fermi} bright blazars reported in \cite{LBAS_SourceList}. The detailed SEDs have been built and studied using simultaneous or quasi-simultaneous data taken from May 2008 to January 2009 and obtained from \emph{Fermi}, \emph{Swift}, and other space based or ground telescopes such as AGILE $\gamma$-ray data, Effelsberg, OVRO, Ratan-600 1-22~GHz, radio, mm, NIR and optical data from the GASP-WEBT collaboration, Mid-infrared VISIR observations.

\section{Blazar classification scheme and Data Sample}
\label{sec:classif}

The results of \emph{Fermi} LAT observations in the first three months of science operations are reported in the first LAT Bright Gamma-Ray Source list \cite{LATSourceList} that includes 132 sources at galactic latitude $|$b$|>$10$^\circ$, with test statistic greater than 100 ($>$ 10 $\sigma$). Among these, a sample of 106 sources indicate high-confidence associations with known AGN and is referred to as the LAT Bright AGN Sample (LBAS) \cite{LBAS_SourceList}, comprising 58 FSRQs, 42 BL Lac objects, 4 blazar of uncertain type and 2 radio-galaxies. 
It is confirmed that the bright $\gamma$-ray sky is dominated by radio-loud AGN. In fact about 90$\%$ of the LBAS sources is associated to AGN listed in radio catalogs (CRATES/CGRaBS, BZCat). Moreover only about one third of the bright \emph{Fermi} AGN were also detected by EGRET.

Blazars represent a subgroup of non-thermal dominated AGN. For a more detailed classification of AGN refer to \cite{SEDpaper}. Blazars are extremely strong and highly variable, core dominated flat or inverted spectrum radio loud AGN whose jet is looked closely to the line of sight of the observer.
Blazars can be further divided  into two subgroups: 

\begin{itemize}
 \item FSRQs showing broad emission lines in their optical spectrum just like normal QSO (Quasi Stellar Objects) and including Flat Spectrum Radio Quasars (FSRQs) and broad line radio galaxies.
\item BL Lacs (or BL Lacertae objects) that do not show any strong or broad line in their optical spectrum and whose radio compactness and broad-band SED are very similar to that of strong lined blazars. 
\end{itemize}

\begin{figure}
\includegraphics[width=75mm]{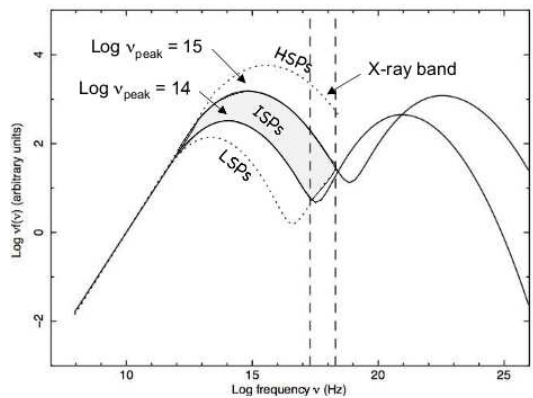}
\caption{Definition of different blazar types based on the peak of the synchrotron component ($\nu_{peak}$) in their SED. Low Synchrotron Peaked blazars, or LSP are those where ($\nu^{S}_{peak}$) is located at frequencies lower than 10$^{14}$~Hz (lower dotted line), Intermediated Synchrotron Peaked sources or ISP 10$^{14} < \nu_{peak} < $10$^{15}$~Hz (SEDs with peak within the gray area).}
\label{fig:classif}
\end{figure}

A further classification of BL Lac objects \cite{PadovaniGiommi} comprises three subclasses (LBL IBL and HBL, as Low, Intermediate and High peaked BL Lacs) depending on the position in the SED of the peak energy of the synchrotron emission that reflects the maximum energy the particles can be accelerated in the jet. This classification has been extended to all blazars that can be classified into Low, Intermediate and High energy (Synchrotron) Peaked blazars, respectively called LSP (synchrotron peak $\nu_{peak} < $10$^{14}$~Hz), ISP (10$^{14} < \nu_{peak} < $10$^{15}$) and HSP ($\nu_{peak} > $10$^{15}$) 
In the LSP case, the X-ray emission is flat and due to the rising part of the Inverse Compton component. At these relatively low energies it occurs in the Thomson regime; in the ISP case, the X-ray band includes both the tail of the synchrotron emission and the rise of the Inverse Compton component. Finally, in the HSP case, where the emitting particles are accelerated at much higher energies than in LSPs, the synchrotron emission dominates the observed flux in the X-ray band and the inverse Compton scattering occurs in the Klein-Nishina regime (see Fig.\ref{fig:classif}).

\section{Radio to $\gamma$-ray SEDs}
The 48 sources chosen from the LBAS list to study their broad band SED using simultaneous or quasi simultaneous radio to high-energy $\gamma$-ray data, were chosed on the basis of the availability of \emph{Swift} observations carried out between May 2008 and January 2009 (which have been scheduled largely independently of \emph{Fermi} results) and not on brightness level or any other condition that could influence the shape of the SED. This was checked by verifying that the distributions of redshift, optical, X-ray and $\gamma$-ray fluxes are all consistent with being the same in both subsamples.
The quasi-simultaneous SEDs have been built in the usual Log$\nu$-Log($\nu$F($\nu$)) representation.
A detailed description of the data analysis performed to derive the SEDs is reported in \cite{CutiniSymp}.
It is worth to underline that all SEDs exhibit the typical double hump distribution, with the first bump attributed to synchrotron radiation and the second one likely due to one or more components related to inverse Compton emission. 

\begin{figure}
\includegraphics[width=75mm]{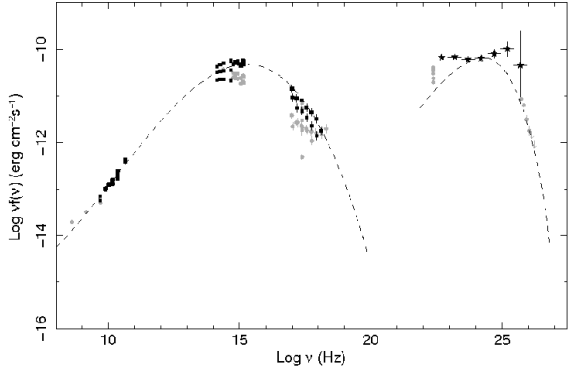}
\caption{The SED of 0FGL J0222.6+4302 = 3C66A . Fermi data appear as black stars; the quasi-simultaneous data at other wavelenghts are represented by filled black squares; the non-simultaneous archival measurements are shown as gray small dots; finally filled black triangles show AGILE data. The dashed lines represent the best fits to the Synchrotron and inverse Compton part of the quasi-simultaneous SEDS  (see \cite{SEDpaper} for details).}
\label{fig:3C66A}
\end{figure}

\begin{figure}
\includegraphics[width=75mm]{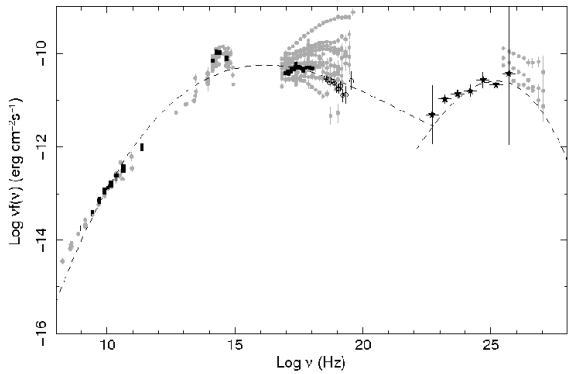}
\caption{The SED of 0FGL J1653.9+3946 = MRK501 . Fermi data appear as black stars; the quasi-simultaneous data at other wavelenghts are represented by filled black squares; the non-simultaneous archival measurements are shown as gray small dots; finally filled black triangles show AGILE data. The dashed lines represent the best fits to the Synchrotron and Inverse Compton part of the quasi-simultaneous SEDS  (see \cite{SEDpaper} for details).}
\label{fig:MRK501}
\end{figure}

Fig.\ref{fig:3C66A} and \ref{fig:MRK501} show the SEDs of two sources among the 48 LBAS: 3C66A and MRK501 which can be classified as an ISP and HSP respectively. 
A comparison between the two SEDs shows the differences in the position of the peaks of the two components and of the relative peak intensities. This behaviour is generally found in all SEDs. Large variability is also present, especially at optical/UV and X-ray frequencies. Gamma-ray variability cannot be evaluated as the \emph{Fermi} data used for our analysis are averaged over the entire period of data taking. The $\gamma$-ray variability of the LBAS blazars is separately discussed in \cite{blazarsvariability}, while a complete description of the $\gamma$-ray spectral shape of the LBAS sources is given in \cite{LBASpaper}. From the results of the latter paper, we note that often \emph{Fermi} data cannot be fit by a simple power law as significant curvature is detected. Convex curvature is often observed in sources where synchrotron peak is located at low energies (e.g. PKS0454-234, PKS1454-354 and PKS1502+106, 3C454.3 etc) whereas very flat or even concave type curvature is exhibited by high synchrotron peaked objects (e.g. 3C66A, see Fig.\ref{fig:3C66A}, PKS0447-439, 1ES~0502+675 and PG~1246+586 ).

\section{Observational Parameters of Blazar SEDs}
The key observational parameters that characterize the SEDs of the LBAS sources have been estimated: the radio spectral index ($\alpha_r$), the peak frequency ($\nu^{S}_{peak}$) and the peak intensity flux of the synchrotron component ($\nu^{S}_{peak}$F($\nu^{S}_{peak}$)), the peak frequency ($\nu^{IC}_{peak}$) and the peak intensity flux of the Inverse Compton component ($\nu^{IC}_{peak}$F($\nu^{IC}_{peak}$))

The radio to microwave spectral slope $\alpha_r$ has been measured by performing a linear regression of all radio flux measurements used for the SEDs, both quasi-simultaneous and historical ones. The results of this evaluation is that the measured slope in our SEDs is quite flat ($< \alpha_{r} > \sim$0) and consistent with being the same in all blazar types.

\begin{figure}
\includegraphics[width=75mm]{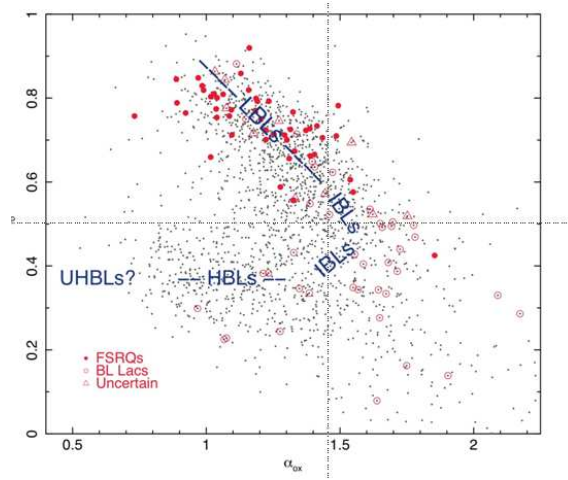}
\caption{The $\alpha_{ox}$-$\alpha_{ro}$ plot of the LBAS blazars sample (large symbols) compared to the blazars from BZCAT catalog which have radio optical and X-ray information (small red symbols).}
\label{fig:alfaxr}
\end{figure}

Fig.\ref{fig:alfaxr} shows the plot of the optical to X-ray spectral slope $\alpha_{ox}$ versus the radio to optical spectral slope $\alpha_{ox}$. The plot includes all blazars in the BZCat catalog for which we have radio, optical and X-ray measurements (small red dots).
\emph{Fermi} FSRQs (filled circles), like all FSRQs discovered in any other energy band, are
exclusively located along the top-left / bottom-right band, whereas BL Lacs (open circles) can be
found in all parts of the plane, even if with a prevalence in the horizontal area defined by values of
$\alpha_{ro}$ between 0.2 and 0.4, which is where HSP sources are located \cite{PadovaniGiommi}.
All $\gamma$-ray selected blazars are located in regions covered by previously known blazars. No new $\gamma$-ray type of blazars has been found, in particular there is no evidence for the hypothetical population of Ultra-High energy peaked blazars (UHBLs), with synchrotron peak in the $\gamma$-ray band ($\nu^{S}_{peak}>$10$^{20}$).

\subsection{Synchrontron peak determination}
\label{sec:Speak}
The peak frequency ($\nu^{S}_{peak}$) and peak intensity $\nu^{S}_{peak}$F($\nu^{S}_{peak}$)  of the synchrotron component of the SEDs have been estimated by fitting the synchrotron dominated part of the SED (from radio to UV in LPS and from radio to X-ray band in HSP blazars) with a simple third degree polynomial function:

\begin{equation}
 \nu F_{\nu} = a\cdot\nu^3+b\cdot\nu^2+c\cdot\nu+d
\end{equation}
\label{polynomialfit}

The synchrotron peak frequency ($\nu^{S}_{peak}$) and the relative peak intensity $\nu^{S}_{peak}$F($\nu^{S}_{peak}$) can be alternatively evaluated through an empirical method that makes use of the spectral slopes $\alpha_{ro}$  between 5~GHz and 5000~$\mathring{A}$ and 
$\alpha_{ox}$  between 5000~$\mathring{A}$ and 1~keV with $\alpha_{ab}$ defined as: 

\begin{equation}
 \alpha_{ab}=-\frac{Log(\frac{f_a}{f_b})}{Log(\frac{\nu_a}{\nu_b})}
\end{equation}

where f$_a$ is the rest-frame flux at frequency $\nu_a$. The value of $\nu^{S}_{peak}$ in the SED of a blazar can be analytically determined from the following relationship:

\begin{equation}
 Log(\nu^{S}_{peak}) = \left\{\begin{array}{ll}
13.85+2.30X & \mbox{if X$<$0 and Y$<$0.3};\\
13.15+6.58Y & \mbox{otherwise}. \end{array}
\right.
\end{equation}

where X=0.565-1.433$\cdot\alpha_{ro}$+0.155$\cdot\alpha_{ox}$ and Y=1.0-0.661$\cdot\alpha_{ro}$-0.339$\cdot\alpha_{ox}$.

This relationship has been calibrated using the $\nu^{S}_{peak}$ values measured from the 48 quasi-simultaneous SEDs and the corresponding $\alpha_{ox}$ and $\alpha_{ro}$ values.

\begin{figure}
\includegraphics[width=75mm]{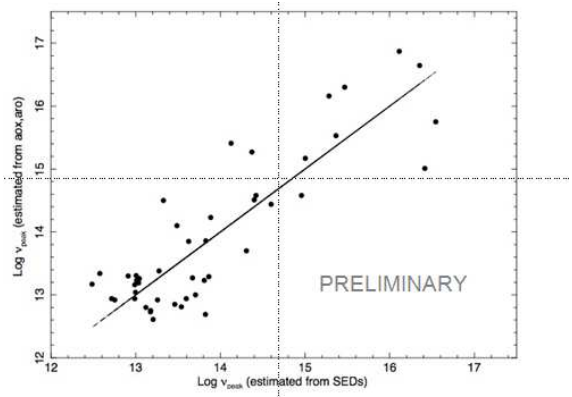}
\caption{The synchrotron peak frequency $\nu_{peak}$ value estimated  using the method based on
$\alpha_{ox}$ and $\alpha_{ro}$ values is plotted against the value estimated from the SEDs. }
\label{fig:peak}
\end{figure}

The results are reported in Fig.\ref{fig:peak}. Despite the fact that $\alpha_ {ox}$ and $\alpha_ {ox}$ are based on non-simultaneous literature data, the distribution of the difference between the values obtained with the two methods has a mean value of 0.04 and a standard deviation of 0.58, implying that the value of Log($\nu^{S}_{peak}$) can be derived even from non-simultaneous values of  $\alpha_{ox}$ and $\alpha_{ro}$ within 0.6 decade at one sigma level and within 1 decade in almost all cases.
It is worth to underline that this method assumes that the optical and X-ray fluxes are not contaminated by thermal emission from the disk or accretion. In blazars where thermal flux components are not negligible, the method described above may lead to a significant overestimation of the position of $\nu^{S}_{peak}$.

\begin{figure}
\includegraphics[width=75mm]{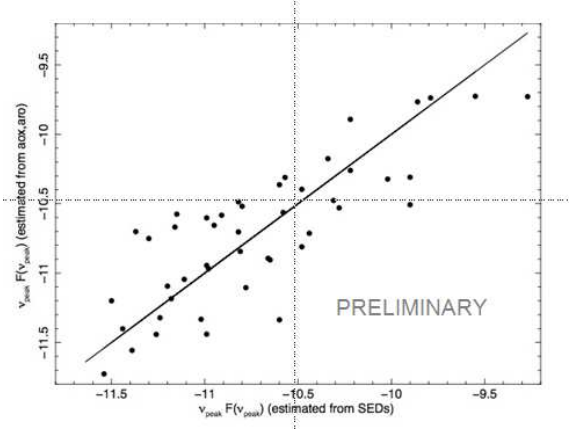}
\caption{The synchrotron peak intensity  $\nu_{peak}$F($\nu^{S}_{peak}$) value estimated  using the method based on
$\alpha_{ox}$ and $\alpha_{ro}$ values is plotted against the value estimated from the SEDs.}
\label{fig:fnu}
\end{figure}

The peak flux $\nu^{S}_{peak}$F($\nu^{S}_{peak}$) can be estimated using the following relationship
\begin{equation}
 Log(\nu_{p}^{S} F(\nu^{S}_{p}) = 0.5\cdot Log(\nu_{p}^{S}) -20.4+0.9\cdot Log(R5GHz)
\end{equation}

where R5GHz is the radio flux density at 5GHz in units of mJy. Fig.\ref{fig:fnu} reports the value of 
$\nu^{S}_{peak}$F($\nu^{S}_{peak}$) estimated with the two methods. Also in this case, the scatter around the solid line, representing perfect match, has an average value of -0.01 for the difference between the two
estimates and a standard deviation of 0.33.
It is quite remarkable that one can derive the synchrotron peak flux simply from $\nu^{S}_{peak}$ and from the radio flux as this implies that within a factor of 10 the radio emission represents a long-term
calorimeter for the whole jet activity and the basic source power.

\subsection{Inverse Compton peak determination}

The peak frequency $\nu^{IC}_{peak}$ and the corresponding peak intensity $\nu^{IC}_{peak}$F($\nu^{IC}_{peak}$) of the inverse Compton bump have been also estimated by fitting the X-ray to $\gamma$-ray part of the SED dominated by inverse Compton emission using the polynomial function of equation \ref{polynomialfit} 
However, for some sources the soft X-ray band is still dominated by synchrotron radiation, and only the Fermi data can be used to constrain the inverse Compton component and the above method is subject to large uncertainties. 
For this reason, in these cases, the simultaneous data points have been fitted to a SSC model with a log-parabolic
electron spectrum \cite{Tramacere} using the ASDC SED \footnote{http://tools.asdc.asi.it/SED/} interface to fit.

The $\nu^{S}_{peak}$F($\nu^{S}_{peak}$) has been determined as the value of $\nu$ that maximizes the $\nu$F($\nu$)
in the IC polynomial function or the predictions of the SSC model. The best fit to both the synchrotron and
inverse Compton components appear as dashed lines in Fig. \ref{fig:3C66A} and \ref{fig:MRK501}.

\begin{figure}
\includegraphics[width=75mm]{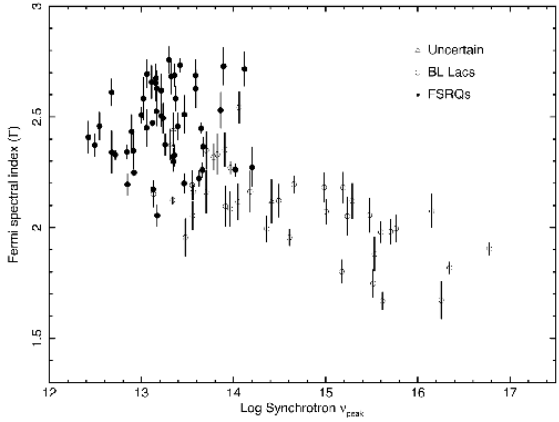}
\caption{The $\gamma$-ray power law photon index ($\Gamma$) plotted against the log of synchrotron peak energy. A clear correlation is present.}
\label{fig:SSCSpeak}
\end{figure}

\begin{figure}
\includegraphics[width=75mm]{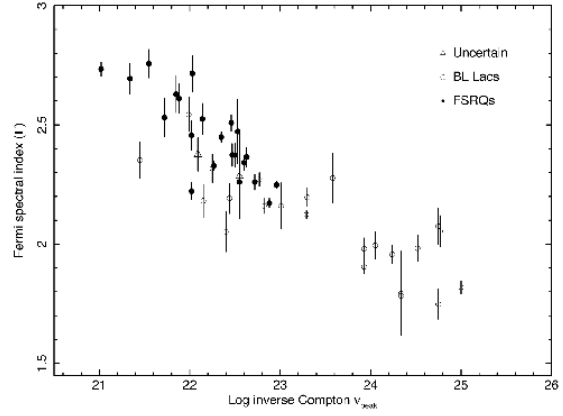}
\caption{The $\gamma$-ray power law photon index ($\Gamma$) plotted against the log of Inverse Compton peak energy. A clear correlation is present.}
\label{fig:SSCICpeak}
\end{figure}

Fig.\ref{fig:SSCSpeak} and \ref{fig:SSCICpeak} respectively show that both the $\nu^{S}_{peak}$ and the $\nu^{IC}_{peak}$ present a strong correlation with their $\gamma$-ray spectral slope $\Gamma$ taken from \cite{LBAS_SourceList}. From the two plots is clear that the scatter is larger in the regions of low $\frac{\nu^{S}_{peak}}{\nu^{IC}_{peak}}$-steep values of $\Gamma$, probably reflecting the presence of $\gamma$-ray spectral slope curvature. The best fit to the $\nu^{IC}_{peak}$-$\Gamma$ relationship is

\begin{equation}
 Log(\nu_{peak}^{IC}) = -4.0 \cdot \Gamma +31.6
\end{equation}
\label{ICpeakfit}

Since the 48 sources for which we have quasi-simultaneous SEDs are representative of the entire LBAS sample, the above equation has been also used to estimate the  $\nu^{IC}_{peak}$ of the LBAS sources for which we have no simultaneous SEDs \cite{SEDpaper}.
The statistical uncertainty associated to $\nu^{IC}_{peak}$ calculated via eq.\ref{ICpeakfit} can be estimated from the distribution of the difference between $\nu^{IC}_{peak}$ measured from the SED and that from eq.\ref{ICpeakfit}. This distribution is centered on the value of 0 and has a sigma of 0.51;
considering that the value of $\nu^{IC}_{peak}$ from the SED is also subject to a a similar error we conservatively conclude that Log($\nu^{IC}_{peak}$) estimated through eq.\ref{ICpeakfit} has an associated error of about 0.7.

\section{Distribution of synchrotron and inverse Compton peak frequencies}
The new classification scheme of blazars introduced in section \ref{sec:classif} and the method used to estimate  
$\nu^{S}_{peak}$, described in section \ref{sec:Speak} allow to classify the LBAS sample in terms of its content of LSP, ISP and HSP objects and to compare it with that of samples selected in other energy bands.
The distribution of the synchrotron peak frequency ($\nu^{S}_{peak}$) of LBAS blazars (estimated using
the $\alpha_{ox}$-$\alpha_{ro}$ method) shows that while the $\nu^{S}_{peak}$ distribution of FSRQs starts at $\sim$10$^{12.5}$~Hz, peaks at 10$^{13.3}$~Hz and it does not extend beyond $\approx$10$^{14.5}$~Hz, the distribution of BL Lacs is much flatter, $\sim$10, it starts at $\sim$10$^{13}$~Hz and reaches much higher frequencies ($\approx$10$^{17}$~Hz) than that of FSRQs. This result has been compared with the distribution of $\nu^{S}_{peak}$ of the sample of FSRQs and BL Lacs detected as foreground sources in the WMAP 3-year microwave anisotropy maps \cite{wmapGiommi}
and with the $\nu^{S}_{peak}$ distribution of the LBAS sample with that of the X-ray selected sample of blazars
detected in the Einstein Extended Medium Sensitivity Survey \cite{emssGioia}.
It can be noted that the $\nu^{S}_{peak}$ distribution of FSRQs is consistent with being the same in the $\gamma$-ray,
radio/microwave and in the X-ray band. Moreover, the large majority of FSRQs are of the LSP type while no FSRQs of the HSP type have been found at any frequency. On the contrary, the $\nu^{S}_{peak}$ distribution of BL Lac objects is very different in the three energy bands. It is strongly peaked at $\sim$10$^{13.3}$~Hz in the microwave band, where HSP sources are very rare, whereas in the X-ray and and $\gamma$-ray bands HSP sources are more abundant than LSPs.
On the other side, the distribution of the inverse Compton peak frequency $\nu^{IC}_{peak}$, of the FSRQ (dot-dashed histogram) and the BL Lacs (solid histogram) in the LBAS sample. The two distributions are quite different with the BL Lacs exhibiting much higher  $\nu^{IC}_{peak}$ values. This is most likely due to the same reason that causes the different $\nu^{S}_{peak}$ distributions in the two blazar subclasses.

\begin{figure}
\includegraphics[width=75mm]{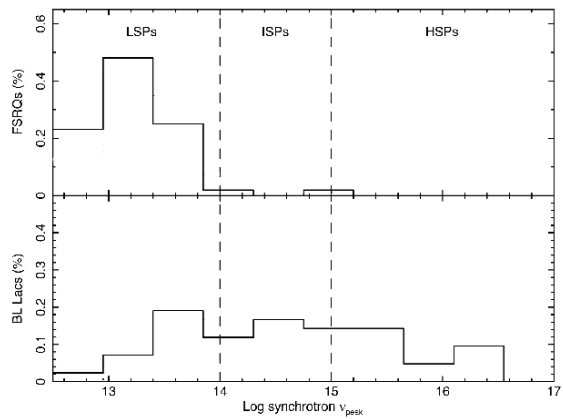}
\caption{The distribution of synchrotron peak energy for the sample of LBAS FSRQ (top panel) and BLLacs (bottom panel).}
\label{fig:Speakdist}
\end{figure}

\begin{figure}
\includegraphics[width=75mm]{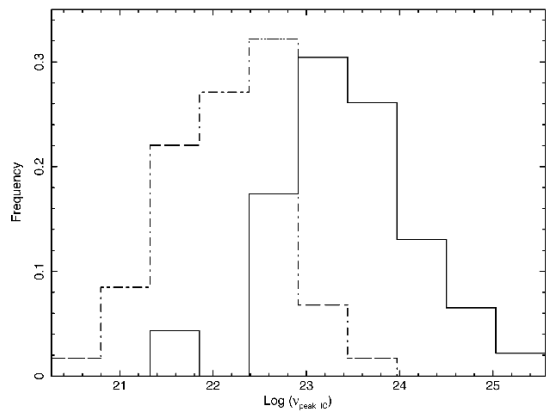}
\caption{The distribution of the inverse Compton peak energy for the sample of LBAS FSRQ (dot-dashed line) and BLLacs (solid line).}
\label{fig:ICpeakdist}
\end{figure}

\section{Conclusions}

A sample of LBAS blazars detected by the \emph{Fermi} LAT in the first three months of scientific operations has been selected to build simultaneous or quasi-simultaneous SEDs using data obtained also from \emph{Swift}, radio/mm telescopes and infra-red and optical facilities.
The collection of SEDs allowed us to estimate important parameters characterizing the SED of $\gamma$-ray selected blazars, such as the frequency and intensity of both the synchrotron and inverse Compton peaks, both directly from the simultaneous data or indirectly using a refined version of the method proposed in \cite{PadovaniGiommi} based on the position in the $\alpha_{ox}$-$\alpha_{ro}$ plane for the former and on the slope of the $\gamma$-ray spectrum for the latter, as the $\gamma$-ray spectral slope and $\nu^{IC}_{peak}$ are strongly correlated.
Among the several results we obtained, we can conclude that $\gamma$-ray selected blazars have
broad-band spectral properties similar to those of radio and X-ray discovered blazars implying that
they are all drawn from the same underlying population; the distribution of the synchrotron peak
frequency is very different for the FSRQ and BL Lac subsamples. The results rule out the existence
of FSRQs of the HSP type, consistent with what also observed in radio, microwave and X-ray surveys; the BL Lac minimum  $\nu^{S}_{peak}$appears to be larger than in FSRQs and this could be due to some intrinsic difference in the mechanism of particle acceleration in the two types of blazars or to a mere selection effect.


\bigskip 
\begin{acknowledgments}

The {\em Fermi} LAT Collaboration acknowledges support from
a number of agencies and institutes for both development and
the operation of the LAT as well as scientific data
analysis. These include NASA and DOE in the United States,
CEA/Irfu and IN2P3/CNRS in France, ASI and INFN in
Italy, MEXT, KEK and JAXA in Japan, and the 
K.~A.~Wallenberg Foundation, the Swedish Research
Council and the National Space Board in Sweden.
Additional support from INAF in Italy for science
analysis during the operation phase is also
gratefully acknowledged.

This research
is based also on observations with the 100 m telescope of the MPIfR
(Max-Planck-Institut f\"{u}r Radioastronomie) at Effelsberg. The OVRO
$40~m$ program is supported in part by NASA (NNX08AW31G) and the NSF
(AST-0808050).

\end{acknowledgments}

\bigskip 


\end{document}